\documentclass [a4paper,fleqn]{article}
\usepackage{graphicx}
\usepackage {subfigure,epsfig}

\usepackage {amsmath} \usepackage{amssymb} \usepackage{cite} \usepackage{amsthm}
\numberwithin{equation}{section} \numberwithin{table}{section} \mathindent=0pt
\theoremstyle{plain} \newtheorem{theorem}{Theorem}

\numberwithin{theorem}{section}

\begin{document}

\title{Simplest equation method to look for exact solutions of nonlinear differential equations}
\author{N.A. Kudryashov}
\date{Department of Applied Mathematics\\
Moscow  Engineering and Physics Institute\\
(State university)\\
31 Kashirskoe Shosse,  115409\\
Moscow, Russian Federation} \maketitle

\begin{abstract}
New method is presented to look for exact solutions of nonlinear
differential equations. Two basic ideas are at the heart of our
approach. One of them is to use the general solutions of the
simplest nonlinear differential equations. Another idea is to take
into consideration all possible singularities of equation studied.
Application of our approach to search exact solutions of nonlinear
differential equations is discussed in detail. The method is used
to look for exact solutions of the Kuramoto - Sivashinsky equation
and the equation for description of nonlinear waves in a
convective fluid. New exact solitary and periodic waves of these
equations are given.
\end{abstract}

\emph{Keywords:} exact solution, nonlinear differential equation,
travelling wave, simplest equation method.\\

PACS: 02.30.Hq - Ordinary differential equations

\section{Introduction}

One of the most exciting recent advances of nonlinear science and
theoretical physics has been a development of methods to look for
exact solutions for nonlinear differential equations. This is
important because many mathematical models are described by
nonlinear differential equations.

The inverse scattering transform \cite{1} and the direct method by
R. Hirota \cite{2} are known as impressive methods to search
solutions of exactly solvable differential equations.

The singular manifold method \cite{3,4,5}, the transformation
method \cite{6,7,8}, the tanh-function method \cite{9,10,11,12},
the sh-function method \cite{13,14} and the Weierstrass function
method \cite{15,16} are useful in many applications to search
exact solutions of nonlinear partially solvable differential
equations.

In this paper we present new approach to look for exact solutions
of nonlinear differential equations. We take two basic ideas for
our method into account. The first idea is to apply the simplest
nonlinear differential equations (the Riccati equation, the
equation for the Jacobi elliptic fuction, the equation for the
Weierstrass ellipic function and so on) that have lesser order
then the equation studied. The second idea is to use all possible
singularities of the equation studied.

There are two advantages of our method. The first advantage is
that our approach generalizes a number of methods that were
applied before. The second advantage of the method is simplicity
of realization.

We apply our approach to look for exact solutions of the Kuramoto
- Sivashinsky equation \cite{17,18,19} and for the equation at the
description of nonlinear waves on the surface in a convective
fluid \cite{20,21}.

The outline of this paper is as follows. New approach to look for
exact solution of nonlinear differential equation is discussed in
section 2. Application of our approach to look for exact solitary
and periodic waves of the Kuramoto - Sivashinsky equation are
given in sections 3 and 4.  Exact solitary and periodic waves of
the equation for nonlinear waves in a convective fluid are
presented in sections 5 and 6.

\section{Method applied}

Consider the nonlinear ordinary differential equation in the
polynomial form

\begin{equation}
M[y]=0\label{e:2.1}
\end{equation}

Let us assume we look for exact solutions of equation
\eqref{e:2.1}. Take the nonlinear ordinary differential equation
of lesser order than equation \eqref{e:2.1}

\begin{equation}\label{e:2.2}
E_n[Y]=0
\end{equation}

 Any nonlinear ordinary differential equation \eqref{e:2.2}
of lesser order than \eqref{e:2.1} with known general solution we
call as the simplest equation.

The first example of the simplest equation is the Riccati equation
\begin{equation}\label{e:2.3}
E_1[Y]=Y_z+Y^2- a Y-b=0
\end{equation}

The second example of the simplest equation  is the equation for
the Jacobi elliptic function
\begin{equation}\label{e:2.4}
E_2[Q]=Q_z^2-Q^4-a Q^3-b Q^2-c Q -d=0
\end{equation}
The third example of the simplest equation is the equation for the
Weierstrass elliptic function
\begin{equation}\label{e:2.5}
E_3[R]=R_z^2-4R^3-a R^2-b R-c =0
\end{equation}

We emphasize that the general solutions of equations \eqref{e:2.3}
and \eqref{e:2.4} have the first order singularities but the
general solution of equation \eqref{e:2.5} has the second order
singularity. We are going to use these properties of equations
\eqref{e:2.3}, \eqref{e:2.4} and \eqref{e:2.5}.

Let us show that the general solution of equation \eqref{e:2.4} is
written via the Jacobi elliptic function sn$(mz, k)$. Equation
\eqref{e:2.4} can be presented in the form

\begin{equation}\label{e:2.6v}
Q_z^2 = (Q-Q_1) (Q-Q_2)(Q-Q_3) (Q-Q_4)
\end{equation}
where $Q_1, Q_2, Q_3$ and $Q_4$ are roots of the equation

\begin{equation}\label{e:2.7v}
Q^4 + a Q^3 + bQ^2 + cQ + d=0
\end{equation}

Taking new variable

\begin{equation}\label{e:2.8v}
Q^2= \frac{(Q_2 - Q_4)(W(z) -Q_1)}{(Q_1-Q_4)(W(z)-Q_2)}
\end{equation}

with

\begin{equation}\label{e:2.9v}
k^2=\frac{(Q_2-Q_3)(Q_1-Q_4)}{(Q_1-Q_3)(Q_2-Q_4)}
\end{equation}

\begin{equation}\label{e:2.10v}
m^2=\frac14{(Q_2-Q_4)(Q_1-Q_3)}
\end{equation}
into consideration we have the equation

\begin{equation}\label{e:2.11v}
\left(\frac{dW}{dz}\right)^2=\left(1-W^2)(1-k^2W^2\right)
\end{equation}
with the general solution

\begin{equation}\label{e:2.12v}
W (z)=\textup{sn} (mz, k)
\end{equation}

The general solution of equation \eqref{e:2.5} is expressed via
the Weierstrass elliptic function because having used new variable

\begin{equation}\label{e:2.13v}
R=\wp -\frac{a^2}{12}
\end{equation}
we obtain the equation for the Weierstrass elliptic function
\cite{22}

\begin{equation}\label{e:2.14v}
\begin{gathered}
\wp^2=4\wp^3-g_2\wp-g_3,\\
\end{gathered}
\end{equation}
where $g_2=\frac{a^2}{12}-b,\,\,\,
g_3=\frac{ab}{12}-c-\frac{a^3}{216}$.

 Suppose we can find the relation between solutions of
 equations \eqref{e:2.1} and \eqref{e:2.2}

\begin{equation}
\label{e:2.6}y=F(Y)
\end{equation}

In addition, suppose that substitution \eqref{e:2.6} into equation
\eqref{e:2.1} leads to the relation in the form

\begin{equation}
\label{e:2.7}M[F(Y)]=\hat{A}E_n[Y]
\end{equation}
where $\hat{A}$ is an operator that should be found.

we can see from relation \eqref{e:2.7} that for any solution of
the simplest equation \eqref{e:2.2} there exists a special
solution of equation \eqref{e:2.1} by the formula \eqref{e:2.6}.

Later we will show that there is the transformation \eqref{e:2.6}
that allows us to obtain relation \eqref{e:2.7}.

Consider our method in detail. It contains the following steps.

\textit{The first step.} Determination of the dominant terms and
the singularity order for the solution of equation \eqref{e:2.1}.

To find the dominant terms of equation \eqref{e:2.1} we substitute

\begin{equation}\label{e:2.8}
y=A_k(z-z_0)^k
\end{equation}
into all terms of equation \eqref{e:2.1}.

It should be noted that the first step of our method correspond to
the first step of the Painleve test that is used to analyze
equation \eqref{e:2.1} on the Painleve property \cite{23}. We have
to compare the degrees of all terms of equation \eqref{e:2.1} and
take two or more of them with the least degree. Let $N$ be the
least degree of the dominant terms of equation \eqref{e:2.1} and
denote by $n$ the value of $k$ that corresponds to the least
degree.

 The value of $k$ that leads to the terms of the least
degree by substitution of \eqref{e:2.8} into equation
\eqref{e:2.1} we call as the singularity order of the solution of
equation \eqref{e:2.1}.

Our method can be applied when $n$ is integer. In the case of non
- integer value of $n$ one can try to transfer equation
\eqref{e:2.1}.

\textit{The second step.} Arrangement of formula \eqref{e:2.6} as
polynomial dependence on the general solution of the simplest
equation \eqref{e:2.2}. This formula can be written taking into
consideration the singularity orders for the solutions of
equations \eqref{e:2.1} and \eqref{e:2.2}.

Taking equation \eqref{e:2.3} into consideration in this paper we
firstly introduce and use new formula \eqref{e:2.6} to look for
exact solutions of nonlinear differential equations that takes the
form

\begin{equation}
\label{e:2.13}y(z)=A_0+A_1 Y+ \ldots +A_n Y^n+B_1
\left(\frac{Y_z}{Y}\right)+ \ldots +B_n
\left(\frac{Y_z}{Y}\right)^n
\end{equation}
where $Y(z)$ is the general solution of equation \eqref{e:2.3},
coefficients $A_k\,\,\,(k=0, \ldots, n)$ and $B_k\,\,\,(k=1,
\ldots, n)$ are  unknown coefficients  that should be found.

It should be noted that assuming $B_k=0\,\,\,(k=1, \ldots, n)$ in
formula \eqref{e:2.13} and $a=0$ in equation \eqref{e:2.3} our
approach is reduced to the tanh-function method to look for exact
solution of nonlinear differential equation \eqref{e:2.1}. For
$a=0$ in equation \eqref{e:2.3} we obtain the method of the
hyperbolic functions from formula \eqref{e:2.13} to look for exact
solutions.

Taking  equation \eqref{e:2.4} into consideration  we also firstly
introduce  and we apply formula \eqref{e:2.6} in the form

\begin{equation}
\begin{gathered}
\label{e:2.14}y(z)=A_0+A_1\,Q + \ldots + A_n \,Q^n + \left(B_1 +
\ldots
+ B_{n-1}\, Q^{n-2}\right) Q_z+\\
\\
+D_1\left(\frac{Q_z}Q\right) + D_2 \left(\frac{Q_z}Q\right)^2
+\ldots+ D_n \left(\frac{Q_z}Q\right)^n,
\end{gathered}
\end{equation}
where $Q(z)$ is the general solution of equation \eqref{e:2.4},
$A_k\,\,\,(k=0, \ldots, n)$, $B_k\,\,\,(k=1,\ldots, n-1)$ and
$D_k\,\,\,(k=1, \ldots, n)$ are coefficients that should be found.
For $B_k=0,\,\,\,(k=0,\ldots, n)$ and $D_k=0\,\,\,(k=1, \ldots,
n)$ formula \eqref{e:2.14} can be reduced to the sn-function
method to look for exact solutions of equation \eqref{e:2.1}.

Taking  equation \eqref{e:2.5} into account  we can also suggest
formula \eqref{e:2.6} to look for exact solutions of nonlinear
differential equations in the form

\begin{equation}
\begin{gathered}
\label{e:2.16}y(z)=A_0 +D_1\left(\frac{R_z}R\right)+ D_2
\left(\frac{R_z} R\right)^2 +\ldots + D_n \left(
\frac{R_z}R\right)^n+ \\
\\
+A_1 R + B_1 R_z + (A_2 R + B_2 R_z) R +(A_3 R + B_3R_z) R^2 +
\ldots,
\end{gathered}
\end{equation}

where $R(z)$ is the general solution of equation \eqref{e:2.5}.
The last term in formula \eqref{e:2.16} is determined by the
singularity order of the solution of equation \eqref{e:2.1}. If
$n=2m$ then the last term is $A_m R^m$. In the case $n=2m-1$ the
last term in formula \eqref{e:2.16} is $B_m R^{m-1} R_z$.

Assuming $A_k=0\,\,\,(k=2,\ldots,n)$, $D_k=0\,\,\,(k=1,\ldots,n)$,
$B_k=0\,\,\,(k=2,\ldots,n)$ from formula \eqref{e:2.16} we have
the approach by N.A. Kudryashov \cite{15} that was used to find
exact solutions of nonlinear differential equation. Using
$A_k=0\,\,\,(k=2,\ldots,n)$, $D_k=0\,\,\,(k=2,\ldots,n)$ and
$B_k=0 \,\,\,(k=1,\ldots,n)$  from expression \eqref{e:2.16} we
obtain the method by A.V. Porubov \cite{16}.

\textit{The third step}. Determination of coefficients
$A_k\,\,\,(k=0,\ldots,n)$, $B_k\,\,\,(k=1,\ldots, n)$ and
$D_k\,\,\,(k=1,\ldots,n)$ in formulas \eqref{e:2.13},
\eqref{e:2.14} and \eqref{e:2.16}.

Using formulas \eqref{e:2.13}, \eqref{e:2.14} or \eqref{e:2.16}
for finding exact solutions of equation \eqref{e:2.1} we chose the
type of solution for equation \eqref{e:2.1}. If we use formula
\eqref{e:2.13} we can find exact solutions of equation
\eqref{e:2.1} in the form of solitary waves. Using formula
\eqref{e:2.14} we can look for special periodic solutions
expressed via the Jacobi elliptic functions. If we use formula
\eqref{e:2.16} we can obtain periodic exact solutions expressed
via the Weierstrass elliptic function.

Coefficients $A_k\,\,\,(k=0,\ldots, n)$, $B_k\,\,\,(k=1, \ldots,
n)$ and $D_k\,\,\,(k=1,\ldots, n)$ are determined after
substitution of formulas \eqref{e:2.13}, \eqref{e:2.14} and
\eqref{e:2.16} into equation \eqref{e:2.1}. As a result of this
substitution we obtain the equation with derivatives of $Y$ (or
$Q$ and $R$) with respect to $z$. To find coefficients $A_k\,\,\,
(k=0, \ldots, n)$, $B_k\,\,\,(k=1,\ldots, n)$ and
$D_k\,\,\,(k=1,\ldots, n)$ we take the following simple theorems
into consideration.

\begin{theorem}
\label{T:2.1.} Let $Y(z)$ be a solution of equation \eqref{e:2.3}.
Than the equations

\begin{equation}
\label{2.18}Y_{zz} =2Y^3 - 3a Y^2 + (a^2 -2b) Y + ab
\end{equation}

\begin{equation}
\begin{gathered}
\label{e:2.19}Y_{{{\it zzz}}}=-6\, Y ^{4}+12\,
  Y ^{3}a+ \left( 8\,b-7\,{a}^{2}\right)
  Y  ^{2}+ \\
\left( -8\,ab+{a}^{ 3} \right) Y -2\,{b}^{2}+{a}^{2}b
\end{gathered}
\end{equation}

\begin{equation}
\begin{gathered}
\label{e:2.20}Y_{{{\it zzzz}}}=24\, Y ^{5}-60\,
  Y ^{4}a+ \left( 50\,{a}^{2}-40\,b
 \right)  Y ^{3}+ \\
 +\left( 60\,ab-15\,{
a}^{3}\right) Y ^{2}+ \left( 16\,{b }^{2}-22\,{a}^{2}b+{a}^{4}
\right) Y -8\,a{b}^{2}+{a} ^{3}b
\end{gathered}
\end{equation}

\begin{equation}
\begin{gathered}
\label{e:2.21}Y_{{{\it zzzzz}}}=-120\,  Y ^{6}+360\,
  Y ^{5}a+ \left( -390\,{a}^{2}+240\,b
 \right)  Y ^{4}+\\
 + \left( 180\,{a}^{3}
-480\,ab \right)  Y ^{3}+ \left( -136
\,{b}^{2}-31\,{a}^{4}+292\,{a}^{2}b \right)  Y ^{2}+ \\
 +\left( 136\,a{b}^{2}-52\,{a}^{3}b+{a}^{5}\right)
 Y -22\,{a}^{2}{b}^{2}+{a}^{4}b+16\,{b}^{3}
\end{gathered}
\end{equation}
have special solutions that are expressed via the general solution
of equation \eqref{e:2.3}.
\end{theorem}

\begin{proof}Theorem 2.1 is proved by differentiation of \eqref{e:2.3} with respect to $z$
and substitution $Y_z$ from equation \eqref{e:2.3} into
expressions obtained.

\end{proof}

\begin{theorem}
\label{T:2.2.} Let $Q(z)$ be a solution of equation \eqref{e:2.4}.
Than the equations

\begin{equation}
\label{e:2.22}Q_{{{\it zz}}}=2\,  Q ^{3}+\frac32\,a
  Q ^{2}+bQ +\frac12\,c
\end{equation}

\begin{equation}
\label{e:2.23}Q_{{{\it zzz}}}=6\,  Q ^{2} Q _{{z}} +3\,aQ \left( z
\right) Q_{{z}}
 +b Q_{{z}}
\end{equation}

\begin{equation}
\begin{gathered}
\label{e:2.24}Q_{{{\it zzzz}}}=24\, Q ^{5}+30\,a
 Q ^{4}+ \left( \frac{15}2\,{a}^{2}+20\,b
 \right)  Q ^{3}+ \\
 +\left( {15}2\,ab+15
\,c \right)  Q ^{2}+ \left( {b}^{2}+\frac92\,ac+12\,d \right) Q
+3\,ad+\frac12\,bc
\end{gathered}
\end{equation}

\begin{equation}
\begin{gathered}
\label{e:2.25}Q_{{{\it zzzzz}}}=120\, Q ^{4}Q_{{z}} +120\,a Q
^{3}Q_{{z}}+3\, \left( \frac{15}2\,{a}^{2}+20\,b
 \right)  Q ^{2}Q_{{z}} +\\+
 2\, \left( \frac{15}2\,ab+15\,c \right) Q \left( z
 \right) Q _{{z}} + \left( {b}^{2}+\frac92\,ac+12
\,d \right) Q_{{z}}
\end{gathered}
\end{equation}
have special solutions that are expressed via the general solution
of equation \eqref{e:2.4}.
\end{theorem}

\begin{proof}Theorem 2.2 is proved along similar lines than theorem 2.1.
\end{proof}

\begin{theorem}
\label{T:2.3.} Let $R(z)$ be a solution of equation \eqref{e:2.5}.
Than the equations

\begin{equation}
\label{e:2.26}R_{{{\it zz}}}=6\, R ^{2}+a R +\frac12\,b
\end{equation}

\begin{equation}
\label{e:2.27}R_{{{\it zzz}}}=12\,R R_{{z}} +a R_{{z}}
\end{equation}

\begin{equation}
\begin{gathered}
\label{e:2.28}R_{{{\it zzzz}}}=120\, R ^{3}+30\,a
 R ^{2}+ \left( {a}^{2}+18\,b
 \right) R +\frac12\,ab+12\,c
\end{gathered}
\end{equation}

\begin{equation}
\begin{gathered}
\label{e:2.29}R_{{{\it zzzzz}}}=360\, R ^{2}R_{{z}} +60\,aR
R_{{z}} + \left( {a}^{2}+18\,b \right) R_{{z}}
\end{gathered}
\end{equation}
have special solutions that are expressed via the general solution
of equation \eqref{e:2.5}.
\end{theorem}

\begin{proof}Theorem 2.3 is proved in a manner like theorem 2.1.

\end{proof}

Substituting dependence $y$ on $Y$ \eqref{e:2.13} into equation
\eqref{e:2.1} and taking theorem 2.1 into account we obtain

\begin{equation}
\begin{gathered}
\label{e:2.30}M[y] =\sum^{2N}_{k=0} P_k (a, b,A_0, \ldots,
A_n,\,B_1, \ldots, B_n) Y^{k-N}
\end{gathered}
\end{equation}

Consider the system of algebraic equations with respect to
$a,b$,\,$A_{k}\,\,(k=0, \ldots, n)$ and $B_{k}\,(k=1, \ldots, n)$

\begin{equation}
\begin{gathered}
\label{e:2.31}P_k(a,b,A_0,\ldots, A_n,\,B_1, \ldots, B_n)=0\,\,
(k=0, \ldots, 2N)
\end{gathered}
\end{equation}

As a consequence of the choice for the dominant terms of equation
\eqref{e:2.1} we have $A_n\neq0$ at the first step of the method.
Assume that there are solutions of the system of equations
\eqref{e:2.31}. If equations \eqref{e:2.31} are satisfied then

\begin{equation}
\begin{gathered}
\label{e:2.32}M[y]=0
\end{gathered}
\end{equation}
and $y$ by formula \eqref{e:2.13} is the solution of equation
\eqref{e:2.1}. Here $Y(z)$ is the general solution of equation
\eqref{e:2.3}.

Substituting \eqref{e:2.14} into equation \eqref{e:2.1} and taking
theorem 2.2 into consideration we have the equality in the form

\begin{equation}
\begin{gathered}
\label{e:2.35}M[y]=\sum^{N}_{k=-N} P_k (a, b, c, d, A_0,
\ldots, A_n,\,B_1, \ldots,B_{n-1},D_1, \ldots,\,D_n,)\,Q^k +\\
+\sum^{N-2}_{k={2-N}} S_k (a, b, c, d, A_0, \ldots, A_n,\,B_1,
\ldots,B_{n-1},D_1, \ldots,\,D_n\,)\,Q^k\,Q_z.
\end{gathered}
\end{equation}

Suppose we have solutions of the systems of equation

\begin{equation}
\begin{gathered}
\label{e:2.36}P_k (a, b, c, d, A_0, \ldots, A_n,B_1,
\ldots,B_{n-1},D_1, \ldots,\,D_n\,)=0,\,\\
\\
(k=0, \ldots, {2N})
\end{gathered}
\end{equation}

\begin{equation}
\begin{gathered}
\label{e:2.37}S_k\,(a, b, c, d, A_0, \ldots, A_n,B_1, \ldots,
B_{n-1},D_1, \ldots,\,D_n\,)=0,\,\,\,\\
\\
(k=1, \ldots, {2N-4})
\end{gathered}
\end{equation}
then $y$ expressed by the formula \eqref{e:2.14} allows us to have
exact solutions of equation \eqref{e:2.1}.

Substituting \eqref{e:2.16} into equation \eqref{e:2.1} and using
 theorem 2.3 we obtain the relation in the form

\begin{equation}
\begin{gathered}
\label{e:2.38}M[y]=\sum^L_{k=-L} P_k(a, b, c, D_1, \ldots,\,D_n,\,
A_0, A_1 \ldots\,B_1, B_2 \ldots\,)R^k +\\
\\
+\sum^M_{k=-M} S_k  (a, b, c, D_1, \ldots,\,D_n,\,A_0, A_1
\ldots\, B_1, B_2 \ldots\,)R^kR_z
\end{gathered}
\end{equation}

If $N=2m$ then we take in \eqref{e:2.38} $L=N$ and $M=N-2$. In the
case $N=2m-1$ we need to take $M=N$ and $L=N-1$. If we have
solutions for coefficients $D_1, \ldots,\,D_n$; $A_0, A_2, \ldots$
and $B_1, B_2, \ldots$ of the system of the equations

\begin{equation}
\begin{gathered}
\label{e:2.39}P_k(a, b, c, D_1, \ldots,\,D_n, A_0, A_1,
\ldots,\,B_1, B_2, \ldots)=0,\,\, k=0, \ldots, 2L
\end{gathered}
\end{equation}

\begin{equation}
\begin{gathered}
\label{e:2.40}S_k(a, b, c,D_1, \ldots,\,D_n,A_0, A_1, \ldots,B_1,
B_2, \ldots)=0,\,\,\, k=0, \ldots, 2M
\end{gathered}
\end{equation}
we obtain exact solution of equation in the form \eqref{e:2.16}
that is expressed via the Weierstrass solution.

In this manner the method applied allows us to look for exact
solutions of equation \eqref{e:2.1}. Let us consider the
application of our method to look for exact solitary and periodic
waves of nonlinear differential equations.

\section{Exact solitary waves of the Kuramoto -- Sivashinsky equation}

Let us apply our approach to look for exact solitary waves of the
Kuramoto -- Sivashinsky equation. This equation has the form

\begin{equation}
\label{1.1}u_t +uu_{{x}}+\alpha\,u_{{{\it
xx}}}+\beta\,u_{xxx}+\gamma\,u_{xxxx}=0
\end{equation}

Nonlinear evolution equation \eqref{1.1} describes nonlinear waves
in active -- dissipative dispersive media. For $\alpha=0$,
$\gamma=0$ and $\beta\neq0$ from equation \eqref{1.1} we have the
famous Korteveg -- de Vries equation. The Causchy problem for this
equation can be solved by the inverse scattering transform
\cite{1}. In the case $\beta=0$, $\gamma=0$ and $\alpha\neq0$ from
equation \eqref{1.1} we get the Burgers equation that can be
linearized by the Cole -- Hopf transformation into the heat
equation \cite{24,25}. At $\alpha\neq0$, $\beta\neq0$ and
$\gamma\neq0$  equation \eqref{1.1} is not integrable because this
one does not pass the Painleve test\cite{23}. However there are
several special solutions of this equation.

Exact solitary waves of equation  at $\alpha\neq0$, $\beta\neq0$
and $\gamma\neq0$ was firstly found in \cite{4} at $\beta=0;\,\,
\beta=\pm4\sqrt{\alpha\gamma};
\beta=\pm12\sqrt{\alpha\gamma/47};\,\,
\beta=\pm16\sqrt{\alpha\gamma/73}$. Later these solutions were
rediscovered more than once. Let us show that using our approach
we can obtain more general solitary waves.

Using new variables

\begin{equation}
\begin{gathered}
\label{1.31}u=u^{'}\alpha\sqrt{\frac{\alpha}{\gamma}}\,\,\quad\,\,x=x^{'}
\sqrt{\frac{\gamma}{\alpha}},\,\,\quad\,\,t=t^{'}{\frac{\gamma}
{\alpha}^{2}},\,\,\quad\,\,\sigma=\frac{\beta}{\sqrt{\alpha\,\gamma}},
\end{gathered}
\end{equation}
(the primes of the variables are omitted) we have equation in the
form

\begin{equation}
\label{1.31a}u_t +uu_{{x}}+u_{{{\it
xx}}}+\sigma\,u_{xxx}+u_{xxxx}=0
\end{equation}

Taking the travelling wave into consideration

\begin{equation}
\label{1.1a}u(x,t)=y(z),\,\,\quad\,\, z=x-C_{0}\,t
\end{equation}
we obtain from equation \eqref{1.31a}

\begin{equation}
\label{1.2}C_{{1}}-C_{{0}}\,y+y_{{z}}+\sigma\,y_{{{\it
zz}}}+\,y_{{{\it zzz}}}+\frac12\,{y}^{2}=0
\end{equation}

The equation of the dominant terms is

\begin{equation}
\begin{gathered}
\label{1.2a}y_{zzz} +\frac12 y^2=0
\end{gathered}
\end{equation}

Substituting \eqref{e:2.8} into equation \eqref{1.2} we can see
that the general solution of equation \eqref{1.2} has the
singularity order equals three and we take $n=3$ in formula
\eqref{e:2.13}. Therefore we will look for exact solution of
equation \eqref{1.2} in the form

\begin{equation}
\label{1.3}y \left( z \right) =A_{{0}}+A_{{1}}Y +A_{{2}} Y
^{2}+A_{3}\,Y^{3}+{\frac {B_{{1}}Y_{{z}}}{Y}}+{\frac {B
_{{2}}{Y_{{z}}}^{2}}{{Y}^{2}}}+{\frac {B
_{{3}}{Y_{{z}}}^{3}}{{Y}^{3}}}
\end{equation}

It should be noted that formula \eqref{1.3} can be written in the
form

\begin{equation}
\begin{gathered}
\label{1.03}y \left( z \right) =A_{{0}}-2 \,B_{{2}}\,b+\left
(A_{{3}}-B_{{3}}\right )\,{Y}^{3}+\left (B_{{2}}+A_{{2}} \right
)\,{Y}^{2}+\\
\\
+\left (A_{{1}}-B_{{1}}+3\,B_{{3}}\,b\right )\,Y+{\frac
{B_{{1}}b-3\,B_{{3}}\,{b}^{2}}{Y}}+{\frac {B_{{2}}\,{b}
^{2}}{{Y}^{2}}}+{\frac {B_{{3}}\,{b}^{3}}{{Y}^{3}}}
\end{gathered}
\end{equation}

Substituting \eqref{1.03} into equation \eqref{1.2} we have

\begin{equation}
\begin{gathered}
\label{1.3a}B_{{3}}^{(1)}=120,\,\,\,\quad\,\,\,B_{{3}}^{(2)}=0
\end{gathered}
\end{equation}

Consider the case: $B_{3}=B_{3}^{(1)}=120$. We obtain

\begin{equation}
\begin{gathered}
\label{1.4a}B_2=-15\,\sigma,\,\quad\,B_{{1}}=-\frac{15}{76}\sigma^{2}+\frac{60}{19}+240\,b
\end{gathered}
\end{equation}

\begin{equation}
\label{1.5}A_{{0}}=C_{{0}}+\frac{7\,\sigma}{76}-{\frac
{13}{608}}\,\sigma^{3}-20\,b\,\sigma
\end{equation}

\begin{equation}
\label{1.5a}A_{{3}}^{(1)}=240,\,\,\,\quad\,\,\,A_{{3}}^{(2)}=120
\end{equation}

Assuming $A_{{3}}=A^{(1)}_{{3}}=240$ we have additionally

\begin{equation}
\label{1.6}A_{2}=0,\,\,\,\quad\,\,\,A_{{1}}=-\frac{15}{38}\,\sigma^{2}+\frac{120}{19}-240\,b
\end{equation}

We also obtain

\begin{equation}
\label{1.7}b_{{1,2}}=-\frac{5}{4864}\,\sigma^{2}+\frac{5}{304}\pm
\frac{1}{4864}\sqrt{549\,\sigma^{4}-3584\,\sigma^{2}+9216}
\end{equation}

Taking equation \eqref{1.7} into account we find two variants
values $\sigma$. The first variant has

\begin{equation}
\begin{gathered}
\label{1.8}\sigma_{1}^{(1)}=0,\,\,\quad\,\,\sigma_{2,3}^{(1)}=\pm\,4
\end{gathered}
\end{equation}
and the second variant takes the form

\begin{equation}
\begin{gathered}
\label{1.8a}\sigma_{1}^{(2)}=0,\,\,\quad\,\,\sigma_{2,3}^{(2)}=\pm\,4,\,\,\quad\,\,
\sigma_{4,5}^{(2)}=\pm\,\frac{12}{\sqrt{47}},\,\,\quad\,\,\sigma_{6,7}^{(2)}=\pm\,\frac{16}{\sqrt{73}}
\end{gathered}
\end{equation}

These values $\sigma$ corresponds to two variants values $b$. The
first variant is

\begin{equation}
\begin{gathered}
\label{1.9}b_{1}^{(1)}=\frac{11}{304},\,\,\quad\,\,b_{2,3}^{(1)}=\frac{1}{16}
\end{gathered}
\end{equation}

and the second variant takes the form

\begin{equation}
\begin{gathered}
\label{1.10}b_{1}^{(2)}=-\frac{1}{304},\,\,\quad\,\,
b_{2,3}^{(2)}=\frac{1}{1168},\,\,\quad\,\,
b_{4,5}^{(2)}=\frac{1}{752},\,\,\quad\,\,b_{6,7}^{(2)}=-\frac{1}{16}
\end{gathered}
\end{equation}

As a result of calculations for these values $\sigma$ and $b$ we
have the following special solutions of equation \eqref{1.2} for
the first variant

\begin{equation}
\begin{gathered}
\label{1.11}y_{1}^{(1)}(z)=C_{{0}}-\frac{45\,\sqrt{11}}{152\,\sqrt{19}}\left(U+U^{{-1}}\right)+
\frac{165\sqrt{11}}{152\,\sqrt{19}}\,\left(U^{{3}}+U^{{-3}}\right),\,\,\quad\,\,\\
\\
U=\tanh{\left(\frac{\sqrt{11}}{4\sqrt{19}}(z-C_{{2}})\right)}
\end{gathered}
\end{equation}

\begin{equation}
\begin{gathered}
\label{1.12}y_{2,3}^{(1)}(z)=C_{{0}}\pm\,\frac{3}{2}-\frac{15}{8}\,\left(U+U^{-1}\right)\mp
\frac{15}{4}\,\left(U^{{2}}+U^{{-2}}\right)+\\
\\
+\frac{15}{8}\,\left(U^{{3}}+U^{{-3}}\right),\,\,\quad\,\,
U=\tanh{\left(\frac{1}{4}(z-C_{{2}})\right)}
\end{gathered}
\end{equation}

Special solutions for the second variant take the form

\begin{equation}
\begin{gathered}
\label{1.13}y_{1}^{(2)}(z)=C_{{0}}-\frac{135}{152\,\sqrt{19}}\left(U-U^{-1}\right)-
\frac{15}{152\,\sqrt{19}}\,\left(U^{{3}}-U^{{-3}}\right),\,\,\quad\,\,\\
\\
U=\tan{\left(\frac{{1}}{4\sqrt{19}}(z-C_{{2}})\right)}
\end{gathered}
\end{equation}

\begin{equation}
\begin{gathered}
\label{1.14}y_{2,3}^{(2)}(z)=C_{{0}}\pm\frac{30}{\,
\sqrt{73}}+\frac{345}{584\,\sqrt{73}}\left(U+U^{-1}\right)\mp
\frac{15}{\,\sqrt{73}}\left(U^{{2}}+U^{{-2}}\right)+\\
\\
+\frac{15}{584\,\sqrt{73}}\,\left(U^{{3}}+U^{{-3}}\right),\,\,\quad\,\,
U=\tanh{\left(\frac{{1}}{4\sqrt{73}}(z-C_{{2}})\right)}
\end{gathered}
\end{equation}

\begin{equation}
\begin{gathered}
\label{1.15}y_{4,5}^{(2)}(z)=C_{{0}}+\frac{225}{376\,\sqrt{47}}\left(U+U^{-1}\right)\mp
\frac{45}{188\,\sqrt{47}}\left(U^{{2}}+U^{{-2}}\right)\pm\\
\\
\pm\frac{45}{94\,\sqrt{47}}+\frac{15}{376\,\sqrt{47}}\,\left(U^{{3}}+U^{{-3}}\right),\,\quad\,
U=\tanh{\left(\frac{{1}}{4\sqrt{47}}(z-C_{{2}})\right)}
\end{gathered}
\end{equation}

\begin{equation}
\begin{gathered}
\label{1.16}y_{6,7}^{(2)}(z)=C_{{0}}\,\mp\frac{7}{2}\,-\frac{15}{8}\left(U-U^{-1}\right)\mp
\frac{15}{4}\left(U^{{2}}+U^{{-2}}\right)\-\\
\\
-\frac{15}{8}\,\left(U^{{3}}-U^{{-3}}\right),\,\,\quad\,\,
U=\tan{\left(\frac{{1}}{4}(z-C_{{2}})\right)}
\end{gathered}
\end{equation}

Constant $C_{{1}}$ in the case of the first variant is determined
by the formulas

\begin{equation}
\begin{gathered}
\label{1.17} C_{{1}}^{(1)}=\frac{1}{2}\,C_{{0}}^{2}-\frac
{4950}{6859},\,\,\,\quad\,\,\,C_{{2,3}}^{(1)}=\frac{1}{2}\,C_{{0}}^{2}-18
\end{gathered}
\end{equation}

Constant $C_{{1}}$ for the second variant is expressed by the
formulas

\begin{equation}
\begin{gathered}
\label{1.18} C_{{1}}^{(2)}=\frac{1}{2}\,C_{{0}}^{2}+\frac
{450}{6859},\,\,\,\quad\,\,\,C_{{2,3}}^{(2)}=\frac{1}{2}\,C_{{0}}^{2}-\frac{4050}{389017}\,\,\quad\,\,\,\\
\\
C_{{4,5}}^{(2)}=\frac{1}{2}\,C_{{0}}^{2}-\frac{1800}{103823},\,\,\,\quad\,\,\,
C_{{6,7}}^{(2)}=\frac{1}{2}\,C_{{0}}^{2}-8\,\,\quad\,\,\,
\end{gathered}
\end{equation}

Assuming $A_{{3}}=A_{{3}}^{(2)}=120$ we have

\begin{equation}
\begin{gathered}
\label{1.19}A_{{2}}=15\,\sigma,\,\quad\,\,\,A_{{1}}=-\frac{15}{76}\,\sigma^{2}+\frac{60}{19}-120\,b
\end{gathered}
\end{equation}

\begin{equation}
\label{1.20}b_{{1}}=\frac{7}{380}-\frac{13}{3040}\sigma^{{2}},\,\,\quad\,\,b_{{2}}=0
\end{equation}

We also obtain six values for $\sigma$

\begin{equation}
\begin{gathered}
\label{1.21}\sigma_{1,2}=\pm\,\frac{16}{\sqrt{73}},\,\,\quad\,\,\sigma_{3,4}=\pm\,\frac{12}{
\sqrt{47}},,\,\,\quad\,\,\sigma_{5,6}=\pm\,\frac{2\sqrt{14}}{\sqrt{13}}
\end{gathered}
\end{equation}

Constant $C_{1}$ takes values

\begin{equation}
\label{1.22}C_{1,2}=\frac{1}{2}C_{{0}}^{2}-\frac
{4050}{389017},\,\,\quad\,\,C_{3,4}=\frac{1}{2}C_{{0}}^{2}-\frac
{1800}{103823},\,\,\quad\,\,C_{5,6}=\frac{1}{2}C_{{0}}^{2}
\end{equation}

Special solutions of the Kuramoto -- Sivashinsky equation can be
written in the form

\begin{equation}
\begin{gathered}
\label{1.23}y_{{1,2}}(z)=C_{{0}}+\frac{15}{\sqrt{73}}\left(
\,5\,U^{-1}\mp\,4\,U^{-2}\,+U^{-3}\pm\,4
\right),\,\,\quad\,\,\\
\\
U=\tanh{\left(\frac{1}{2\,\sqrt{73}}(z-C_{2})\right)}
\end{gathered}
\end{equation}

\begin{equation}
\begin{gathered}
\label{1.24}y_{{3,4}}(z)=C_{{0}}+\frac{15}{\sqrt{47}}\left(3\,U^{-1}\,\mp
\,3\,U^{-2}+\,U^{-3}\pm3
\right),\,\,\quad\,\,\\
\\
U=\tanh{\left(\frac{1}{2\,\sqrt{47}}(z-C_{2})\right)}
\end{gathered}
\end{equation}

For $\sigma=\sigma_{5,6}$ we have trivial solutions that are
constants.

In the case $B_{{3}}=0$ we obtain

\begin{equation}
\begin{gathered}
\label{1.25}B_{{2}}=0,\,\,\quad\,\,B_{{1}}=0,\,\,\quad\,\,A_{{3}}^{(1)}=120,\,\,\quad\,\,A_{{3}}^{(2)}=0
\end{gathered}
\end{equation}

Assuming $A_{3}=A_{3}^{(1)}=120$ we have

\begin{equation}
\begin{gathered}
\label{1.26}A_{{2}}=-15\,\sigma,\,\,\quad\,\,A_{{1}}=\frac{60}{19}-\frac{15}{76}\,\sigma^{2}-120\,b\,\sigma
\end{gathered}
\end{equation}

\begin{equation}
\begin{gathered}
\label{1.27}A_{{0}}=C_{{0}}+{\frac
{7}{76}}\,\sigma+10\,\sigma\,b-{\frac {13}{608}}\,{ \sigma}^{3}
\end{gathered}
\end{equation}

\begin{equation}
\begin{gathered}
\label{1.28}b_{1,2}={\frac {5}{76}}-{\frac
{5}{1216}}\,{\sigma}^{2}\pm{\frac {1}{1216}}\, \sqrt
{9216-3584\,{\sigma}^{2}+549\,{\sigma}^{4}}
\end{gathered}
\end{equation}

There are two variants values of $\sigma$ corresponding to
$b_{1,2}$:

\begin{equation}
\begin{gathered}
\label{1.29}\sigma_{1}^{(1)}=0,\,\,\quad\,\,\sigma_{2,3}^{(1)}=\pm\,4
\end{gathered}
\end{equation}
and

\begin{equation}
\begin{gathered}
\label{1.30}\sigma_{1}^{(2)}=0,\,\,\quad\,\,\sigma_{2,3}^{(2)}=\pm\,\frac{16}{\sqrt{73}},,\,\,\quad\,\,
\sigma_{4,5}^{(2)}=\pm\,\frac{12}{\sqrt{47}},,\,\,\quad\,\,\sigma_{6,7}^{(2)}=\pm\,4
\end{gathered}
\end{equation}

Special solutions of the Kuramoto -- Sivashinsky equation in the
case of the first variant take the form

\begin{equation}
\begin{gathered}
\label{1.31}y_{1}^{(1)}(z)=C_{{0}}+\frac{15\,\sqrt{11}}{\sqrt{19}}\left(9\,U-11\,U^{{3}}\right),\quad
U=\tanh{\left(\frac{\sqrt{11}}{2\sqrt{19}}(z-C_{{2}})\right)}
\end{gathered}
\end{equation}

\begin{equation}
\begin{gathered}
\label{1.32}y_{2,3}^{(1)}(z)=C_{{0}}\pm\,9+15\left(U\,\mp\,U^{{2}}-U^{{3}}\right),\quad\,
U=\tanh{\left(\frac{1}{2}(z-C_{{2}})\right)}
\end{gathered}
\end{equation}

Constant $C_{1}$ in equation \eqref{1.2} for the special solutions
$y_{{i}}^{(1)}\,(i=1, \ldots,3)$ is given by formula \eqref{1.17}.

Special solutions of the second variant are determined by the
formulas

\begin{equation}
\begin{gathered}
\label{1.32a}y_{1}^{(2)}(z)=C_{{0}}+\frac{15}{\sqrt{19}}\left(3\,U+U^{3}\right),\,\quad\,
U=\tan{\left(\frac{1}{2\sqrt{19}}(z-C_{{2}})\right)}
\end{gathered}
\end{equation}

\begin{equation}
\begin{gathered}
\label{1.32b}y_{2,3}^{(2)}(z)=C_{{0}}-\frac{15}{\sqrt{73}}\left(5\,U\pm\,4U^{2}+U^{3}\mp4\right),\,\,\quad\,\,\\
\\
U=\tanh{\left(\frac{1}{2\sqrt{73}}(z-C_{{2}})\right)}
\end{gathered}
\end{equation}

\begin{equation}
\begin{gathered}
\label{1.33}y_{4,5}^{(2)}(z)=C_{{0}}-\frac{15}{\sqrt{47}}\left(3\,U\pm\,3U^{2}+U^{3}\mp3\right),\,\,\quad\,\,\\
\\
U=\tanh{\left(\frac{1}{2\sqrt{47}}(z-C_{{2}})\right)}
\end{gathered}
\end{equation}

\begin{equation}
\begin{gathered}
\label{1.34}y_{6,7}^{(2)}(z)=C_{{0}}\mp11+15\left(U\mp\,U^{2}+U^{3}\right),\,\,\quad\,\,
U=\tan{\left(\frac{1}{2}(z-C_{{2}})\right)}
\end{gathered}
\end{equation}

Constant $C_{1}$ in equation \eqref{1.2} for the special solutions
$y_{{j}}^{(2)}\,(j=1, \ldots,7)$ is given by formula \eqref{1.18}.

The special solutions \eqref{1.31} - \eqref{1.34} of the Kuramoto
- Sivashinsky equation  were found in \cite{4,10}. The special
solutions of the Kuramoto -- Sivashinsky equation \eqref{1.1} in
the form of the solitary waves \eqref{1.11}, \eqref{1.12},
\eqref{1.13}, \eqref{1.14}, \eqref{1.15}, \eqref{1.16},
\eqref{1.23} and \eqref{1.24} are new.

\section{Exact periodic waves of the Kuramoto - Sivashinsky equation}

Periodic exact solutions of the Kuramoto -- Sivashinsky equation
\eqref{1.1} were firstly obtained at
$\beta=\pm\,4\sqrt{\alpha\gamma}$ taking the Weierstrass elliptic
function into account in \cite{15}. Let us show that our method
allows to look for exact solutions that are expressed via the
Jacobi elliptic functions.

Consider the equation

\begin{equation}
\begin{gathered}
\label{e:3.13}\gamma\,y_{{zzz}} +\beta\,y_{{zz}} +\alpha\,y_{{z}}
 +\frac12\, y ^{2}-C_{{0}}y +C_{{1}}=0.
\end{gathered}
\end{equation}

This equation can be found from equation \eqref{1.1} if we use
travelling wave \eqref{1.1a}.

Let us look for exact solutions of equation \eqref{e:3.13} in the
form

\begin{equation}
\begin{gathered}
\label{e:3.16}y \left( z \right) =A_{{0}}+A_{{1}}Q +A_{{2}}\,
Q^{2}+A_{{3}}\, Q ^{3}+B_{{1}} Q_{{z}}+B_{{2}}Q\,Q _{{z}}+\\
 \\
 +D_1\,\frac{Q_{z}}{Q}+D_2\,\frac{Q_{z}^{2}}{Q^{2}}+D_3\,\frac{Q_{z}^{3}}{Q^{3}},
\end{gathered}
\end{equation}
where $Q(z)$ is solution of equation \eqref{e:2.4}. There is the
case $D_{3}=0$, $D_{2}=0$, $D_{1}=0$ of special solution and we
consider this variant of formula \eqref{e:3.16}. Substituting
\eqref{e:3.16} into equation \eqref{e:3.13} and taking the theorem
2.2 into consideration we have

\begin{equation}
\begin{gathered}
\label{e:3.17}B_{{2}}=-60\,\gamma,\,\,\quad\,A_{{3}}=\pm\,60\,\gamma\,
\end{gathered}
\end{equation}

\begin{equation}
\begin{gathered}
\label{e:3.19}B_{{1}}=\pm\frac{15}{2}\,\beta-15\,\gamma\,a,\,\,\quad\,\,
A_{{2}}=\pm\,45\,\gamma\,a-\frac{15}{2}\,\beta
\end{gathered}
\end{equation}

\begin{equation}
\begin{gathered}
\label{e:3.21}A_{{1}}=-{\frac
{15}{4}}\,\beta\,a\pm\,30\,\gamma\,b\pm\,\frac{10}{3}\,\alpha\mp\,{\frac
{ 5}{24}}\,{\frac {{\beta}^{2}}{\gamma}}
\end{gathered}
\end{equation}

\begin{equation}
\begin{gathered}
\label{e:3.22}A_{{0}}=C_{{0}}-\frac{5}{4}\,\beta\,b+\frac{1}{6}\,{\frac
{\alpha\,\beta}{\gamma}}\pm\,\frac{5}{6}\,
\alpha\,a\pm\,15\,\gamma\,c\mp\,{\frac {5}{96}}\,{\frac
{{\beta}^{2}a}{\gamma} }-{\frac {5}{192}}\,{\frac
{{\beta}^{3}}{{\gamma}^{2}}}
\end{gathered}
\end{equation}

\begin{equation}
\begin{gathered}
\label{e:3.23}d=\frac{1}{4}\,ac-\frac{1}{12}\,{b}^{2}+\frac{1}{12}\,{\frac
{{\alpha}^{2}}{{\gamma}^{2}}}
\end{gathered}
\end{equation}

\begin{equation}
\begin{gathered}
\label{e:3.25}C_{{1}}=15\,b{\alpha}^{2}+\frac{1}{2}\,{C_{{0}}}^{2}-13\,{\frac
{{\alpha}^{3}} {\gamma}}-20\,{\gamma}^{2}\,{b}^{3}+{\frac
{135}{2}}\,{\gamma}^{2}\,b\,c\,a-{ \frac
{135}{2}}\,{\gamma}^{2}\,{c}^{2}+\\
\\
+{\frac {45}{8}}\,{\gamma}^{2}\,{a}^ {2}\,{b}^{2}-{\frac
{45}{8}}\,{a}^{2}\,{\alpha}^{2}-{\frac {135}{8}}\,{
\gamma}^{2}\,{a}^{3}\,c
\end{gathered}
\end{equation}

There is the additional condition for solution of equation
\eqref{e:3.13}: $\beta=\pm4\sqrt{\alpha\gamma}$.

Exact solutions of equation \eqref{e:3.13} at
$\beta=\sqrt{\alpha\,\gamma}$ in the form of the periodic wave
take the form

\begin{equation}
\begin{gathered}
\label{e:3.26}y \left( z
\right)=C_{{0}}\pm\,15\,\gamma\,c-5\,b\,\sqrt
{\alpha\,\gamma}-{\frac {{\alpha}^{3/2}}{\sqrt {
\gamma}}}-15\,a\,\sqrt {\alpha\,\gamma}\,Q\pm\\
\\
\pm\, 30\,b\,\gamma\,Q-30\,\sqrt
{\alpha\,\gamma}\,Q^{2}\pm\,45\,a\,\gamma\,Q^{2}\pm\,60\,\gamma\,Q^{3}-15\,a\,\gamma\,Q_{{z}}\pm\\
\\
\pm\,30\, \sqrt {\alpha\,\gamma}\,Q_{{z}}-60\, \gamma\,Q\,Q_{{z}}
 \end{gathered}
\end{equation}

Here $Q(z)$ is the general solution of the equation for the Jacobi
elliptic function in the form

\begin{equation}
\begin{gathered}
\label{e:3.26a}Q_{z}^{2}-Q^4-a\,Q^3-b\,Q^2-c\,Q-
\frac{1}{12}\,\gamma^2\,\alpha^2+\frac{1}{12}\,b^2-\frac{1}{4}\,a\,c=0
 \end{gathered}
\end{equation}

There is exact solutions of equation \eqref{e:3.13} at
$\alpha=\beta=0$ in the form

\begin{equation}
\begin{gathered}
\label{e:3.27}y \left( z
\right)=C_{{0}}\pm\,15\,\gamma\,c\pm\,30\,b\,\gamma\,Q\pm\,45\,a\,\gamma\,Q^{2}\pm\,60\,\gamma\,Q^{3}-\\
\\
-15\,a\,\gamma\,Q_{{z}}-60\,\gamma\,Q\,Q_{{z}}
\end{gathered}
\end{equation}

Special solutions in the form of periodic waves \eqref{e:3.26} and
\eqref{e:3.27} of the Kuramoto - Sivashinsky equation are
expressed via the Jacobi elliptic function are new.

\section{Exact solitary waves of the equation in a convective fluid}

The equation

\begin{equation}
\begin{gathered}
\label{e:3.28}\gamma y_{zzz} + \beta y_{zz} + \alpha y_z +
\,\frac12\,\varepsilon y^2 + \chi\,yy_z - C_0 y + C_1 =0
\end{gathered}
\end{equation}

can be obtained from the nonlinear evolution equation

\begin{equation}
\begin{gathered}
\label{e:3.29}u_t +\varepsilon\,uu_{x} + \chi\,(uu_x)_x + \alpha
u_{xx} +\beta u_{xxx} + \gamma u_{xxxx}=0
\end{gathered}
\end{equation}
that occurs at the description of nonlinear waves in a convecting
fluid \cite{20}. Assuming $\chi\,=0 $ in \eqref{e:3.28} we have
the Kuramoto - Sivashinsky equation and below we suppose $\chi=1$.

Equation \eqref{e:3.28} does not pass the Painleve test and this
one is not exactly solvable equation but it has several special
solutions that were found in \cite{9,16}.

Let us consider equation \eqref{e:3.28} using our method to look
for new exact solitary waves. Equation for the dominant terms has
the form

\begin{equation}
\begin{gathered}
\label{e:3.30}\gamma y_{zzz} +  yy_z=0
\end{gathered}
\end{equation}

The singularity order for solution of equation \eqref{e:3.28} is
equal to $n=2$ and we can look for exact solution in the form

\begin{equation}
\begin{gathered}
\label{e:3.31}y \left( z \right) =A_{{0}}+A_{{1}}\,Y
+A_{{2}}\,Y^{2}+B_{1}\,\frac{Y_{z}}{Y}+B_{{2}}\,\frac{Y_{z}^{2}}{Y^{2}},
\end{gathered}
\end{equation}
where $Y(z)$ is the general solution of the Riccati equation
\eqref{e:2.3}. Substituting expression \eqref{e:3.31} into
equation \eqref{e:3.28} we have the following values of
coefficients:

\begin{equation}
\begin{gathered}
\label{e:5.1}B_{{2}}^{(1)}=-12\,\gamma,\,\,\quad\,B_{{2}}^{({2})}=0.
\end{gathered}
\end{equation}

Considering the first case $B_{2}=B_{{2}}^{(1)}=-12\,\gamma$ we
obtain

\begin{equation}
\begin{gathered}
\label{e:5.2}A_{{2}}^{(1)}=12\,\gamma,\,\,\quad\,A_{{2}}^{{(2)}}=0
\end{gathered}
\end{equation}

Take $A_{2}=A_{{2}}^{(1)}=12\,\gamma$. In this case we have

\begin{equation}
\begin{gathered}
\label{e:5.3a}B_{{1}}={\frac {12}{5}}\,\beta+24\,\gamma\,a-{\frac
{12}{5}}\,\varepsilon\,\gamma
\end{gathered}
\end{equation}

\begin{equation}
\begin{gathered}
\label{e:5.3b}A_{{1}}=-24\,\gamma\,a+{\frac {12}{5}}\,\beta-{\frac
{12}{5}}\, \varepsilon\,\gamma
\end{gathered}
\end{equation}

\begin{equation}
\begin{gathered}
\label{e:5.3c}A_{{0}}={\frac {23}{25}}\,\beta\,\varepsilon+{\frac
{12}{5}}\,\gamma\,a \varepsilon-{\frac
{24}{25}}\,\gamma\,{\varepsilon}^{2}-4\,\gamma\,{a}^{2}+\frac{1}
{25}\,{\frac {{\beta}^{2}}{\gamma}}-\\
\\
-{\frac
{12}{5}}\,\beta\,a-\alpha-16\,\gamma\,b\\
\end{gathered}
\end{equation}

\begin{equation}
\begin{gathered}\label{e:5.3d}
C_{{0}}={\frac { 1}{125}}\,{\frac
{{\beta}^{3}}{{\gamma}^{2}}}-{\frac
{126}{125}}\,\gamma\,{\varepsilon}^{3}+\frac{4}{5}\,\gamma\,
\varepsilon\,{a}^{2}-\frac{4}{5}\,\beta\,{a}^{2}-{\frac
{3}{125}}\,{\frac { \varepsilon\,{\beta}^{2}}{\gamma}}+{\frac
{128}{125}}\,{\varepsilon}^{2}
\beta-\\
\\
-\alpha\,\varepsilon+\frac{4}{5}\,\gamma\,\varepsilon\,b-\frac{4}{5}\,\beta\,b
\end{gathered}
\end{equation}

\begin{equation}
\begin{gathered}
\label{e:5.4}b^{(1)}={\frac {1}{100}}\,{\frac
{{\beta}^{2}}{{\gamma}^{2}}}-{\frac {1}{50} }\,{\frac
{\beta\,\varepsilon}{\gamma}}+{\frac {1}{100}}\,{\varepsilon}^{2}-
{a}^{2},\,\quad\,\,b^{(2)}=0
\end{gathered}
\end{equation}

\begin{equation}
\begin{gathered}
\label{e:5.5}C_{{1}}={\frac
{409}{1250}}\,{\beta}^{2}{\varepsilon}^{3}-{\frac
{18}{625}}\,{\frac {{\beta}^{4}\varepsilon}{{\gamma}^{2}} }+{\frac
{72}{625}}\,{\frac {{\beta}^{3}{\varepsilon}^{2}}{\gamma}}-{\frac
{553}{625}}\, \gamma\,\beta\,{\varepsilon}^{4}+\\
\\
+\frac{1}{2}\,\varepsilon\,{\alpha}^{2}+
\gamma\,\alpha\,{\varepsilon}^{3}-\alpha\,{\varepsilon}^{2}\beta+{\frac
{589}{1250}}\,{\gamma}^{2}{ \varepsilon}^{5}.
\end{gathered}
\end{equation}

Solution of equation \eqref{e:3.28} in this case has the form

\begin{equation}
\begin{gathered}
\label{e:5.6}y \left( z \right) ={ \frac
{12}{5}}\,\beta\,a-12\,\gamma\,{a}^{2}-{\frac
{22}{25}}\,\gamma\,{\varepsilon}^{2}-\alpha-{\frac {12}{5}}\,\gamma\,a\varepsilon+\\
\\
+{\frac {3} {25}}\,{\frac {{\beta}^{2}}{\gamma}}+{\frac
{19}{25}}\,\beta\,\varepsilon +{\frac {3}{125}}\,{\frac {\left
(\beta-\varepsilon\,\gamma+10\,\gamma\,a \right )\left
(\beta-\varepsilon\,\gamma-10\,\gamma\,a\right )^{2}}{{
\gamma}^{2}\,Y}}-\\
\\
-\frac {3}{2500}\,\frac {\left (\beta-\varepsilon\,
\gamma-10\,\gamma\,a\right )^{2}\left
(\beta-\varepsilon\,\gamma+10\, \gamma\,a\right
)^{2}}{\gamma^{3}\,Y^{2}},
\end{gathered}
\end{equation}
where $Y(z)$ takes the form

\begin{equation}
\begin{gathered}
\label{e:5.7}Y(z)=a+\frac{1}{10}\,\left({\frac
{\beta}{\gamma}}-\epsilon\,\right)\tanh\left(\frac{1}{10}
\,\left({\frac {\beta}{\gamma}}-\epsilon \right)\left (z-C_{{2}}
\right )\right)
\end{gathered}
\end{equation}

We found new exact solutions \eqref{e:5.6} in the form of the
solitary wave with two arbitrary constants $a$ and $C_{2}$ that
satisfies equation \eqref{e:3.28} in the form

\begin{equation}
\begin{gathered}
\label{e:5.8}\gamma\,y_{{{\it zzz}}}+\beta\,y_{{{\it
zz}}}+\alpha\,y_{{z}}+yy_{{z}}
+\frac{1}{2}\,\varepsilon\,{y}^{2}-\frac{4}{5}\left
(\,\gamma\,\varepsilon\,b-\,\beta\,{a}^
{2}-\,\beta\,b\,\right)y-\\
\\
-\left({\gamma\,\varepsilon\,{a}^{2}+\frac {3}{125}}\,{\frac {
\varepsilon\,{\beta}^{2}}{\gamma}}+{\frac
{128}{125}}\,{\varepsilon}^{2} \beta-\alpha\,\varepsilon-{\frac
{126}{125}}\, \gamma\,{\varepsilon}^{3}+{\frac {1}{125}}\,{\frac
{{\beta}^{3}}{{\gamma}^{2}}}\right
)y+\\
\\
+{\frac {589}{1250}}\,{\gamma}^{2}{\epsilon}^{5}-{\frac
{18}{625}}\,{ \frac {{\beta}^{4}\varepsilon}{{\gamma}^{2}}}+{\frac
{72}{625}}\,{\frac { {\beta}^{3}{\varepsilon}^{2}}{\gamma}}+{\frac
{409}{1250}}\,{\beta}^{2}{ \varepsilon}^{3}-{\frac
{553}{625}}\,\gamma\,\beta\,{\varepsilon}^{4}+\\
\\
+\frac{1}{2}\,\varepsilon\,{\alpha}^{2}+\gamma\,{
\varepsilon}^{3}\alpha-{\varepsilon}^{2}\beta\,\alpha=0.
\end{gathered}
\end{equation}

Assuming $\alpha=1$, $\beta=1$, $\gamma=1$ and $\varepsilon=2$ we
have equation \eqref{e:5.8} in simple form

\begin{equation}
\begin{gathered}
\label{e:5.8a}y_{{{\it zzz}}}+y_{{{\it
zz}}}+y_{{z}}+yy_{{z}}+{y}^{2}-{\frac {16}{5}}\,b\,y+{\frac
{751}{125}}\,y+{\frac {10589}{625}} =0
\end{gathered}
\end{equation}

In the case $B_{{2}}=B_{{2}}^{(1)}=-12\,\gamma\,$ and
$A_{{2}}^{(2)}=0$ we have $a=0$ and

\begin{equation}
\begin{gathered}
\label{e:5.9}B_{{1}}=\frac{12}{5}(\beta-\varepsilon\,\gamma),\,\,\quad\,
\,A_{{1}}=\frac{24}{5}(\beta-\varepsilon\,\gamma)
\end{gathered}
\end{equation}

\begin{equation}
\begin{gathered}
\label{e:5.11}A_{{0}}=-{\frac
{24}{25}}\,\gamma{\varepsilon}^{2}-16\,\gamma
b+\frac{1}{25}\,{\frac {{\beta }^{2}}{\gamma}}-\alpha+{\frac
{23}{25}}\,\varepsilon\,\beta
\end{gathered}
\end{equation}

\begin{equation}
\begin{gathered}
\label{e:5.12}C_{{0}}={\frac
{16}{5}}\,\gamma\,\varepsilon\,b-{\frac {16}{5}}\,\beta\,b-{ \frac
{126}{125}}\,\gamma\,{\varepsilon}^{3}-{\frac {3}{125}}\,{\frac
{{\beta}^{ 2}\varepsilon}{\gamma}}+{\frac {1}{125}}\,{\frac
{{\beta}^{3}}{{\gamma}^{2}}}-\\
\\
- \alpha\,\varepsilon+{\frac {128}{125}}\,{\varepsilon}^{2}\beta
\end{gathered}
\end{equation}

\begin{equation}
\begin{gathered}
\label{e:5.13}C_{{1}}={\frac {72}{625}}\,{\frac
{{\varepsilon}^{2}{ \beta}^{3}}{\gamma}}+{\frac
{589}{1250}}\,{\gamma}^{2}\,{\varepsilon}^{5}-{\frac
{553}{625}}\,\gamma\,{\varepsilon}^{ 4}\beta+{\frac
{409}{1250}}\,{\varepsilon}^{3}{\beta}^{2}-{\frac {18}{625
}}\,{\frac
{{\beta}^{4}\varepsilon}{{\gamma}^{2}}}+\\
\\
+\gamma\,\alpha\,{\varepsilon}^{3}+\frac{1}{2}
\,\varepsilon\,{\alpha}^{2}-\alpha\,\,{\varepsilon}^{2}\beta
\end{gathered}
\end{equation}

We have the exact solitary wave in the form

\begin{equation}
\begin{gathered}
\label{e:5.14}y(z)={\frac {3}{50}}\,{\frac {{\beta}^{2
}}{\gamma\,}}-\alpha-{\frac
{47}{50}}\,\gamma{\varepsilon}^{2}+{\frac {22}{25}}\,
\varepsilon\,\beta-12\,\gamma\,Y^{2}\\
\\
+{\frac {3}{500}}\,{\frac {\left (\beta-\varepsilon\,\gamma \right
)^{3}}{{\gamma}^{2}Y}}-{\frac {3}{40000}}\,{\frac {\left (\beta-
\varepsilon\,\gamma\right )^{4}}{{g}^{3}Y^{2}}}+{\frac
{12}{5}}\left (\beta-\,\varepsilon\,\gamma\,\right )Y
\end{gathered}
\end{equation}
where $Y(z)$ satisfies the equation

\begin{equation}
\begin{gathered}
\label{e:5.15}Y_{{z}}+{Y}^{2}-{\frac
{1}{400}}\,{\varepsilon}^{2}+{\frac {1}{200}}\,{ \frac
{\varepsilon\,\beta}{\gamma}}-{\frac {1}{400}}\,{\frac
{{\beta}^{2}}{{\gamma} ^{2}}}=0
\end{gathered}
\end{equation}
and takes the form

\begin{equation}
\begin{gathered}
\label{e:5.16}Y(z)=\frac{1}{20}\,\left ({\frac
{\beta}{\gamma}}-\varepsilon\right )\tanh\left(\frac{1}{20}\,
\left ({\frac {\beta}{\gamma}}-\epsilon\right )\left
(z-C_{{2}}\right )\right).
\end{gathered}
\end{equation}

 Exact solution \eqref{e:3.28} is the special solution of the
 following equation

\begin{equation}
\begin{gathered}
\label{e:5.17}\gamma\,y_{{{\it zzz}}}+\beta \,y_{{{\it
zz}}}+\alpha\,y_{{z}}+yy_{{z}}+\frac{1}{2}\,\varepsilon\,{y}^{2}-\alpha\,\epsilon\,y+\frac{1}{2}\,
\varepsilon\,{\alpha}^{2}+\alpha\,{\varepsilon}^{2}\beta +\gamma\,\alpha\,{\varepsilon}^{3}-\\
\\
-\left ({\frac { 16}{5}}\,\gamma\,\varepsilon\,b-{\frac
{16}{5}}\,\beta\,b-{\frac {126}{125}}\,\gamma{
\varepsilon}^{3}-{\frac {3}{125}}\,{\frac
{{\beta}^{2}\varepsilon}{\gamma}}+{ \frac {1}{125}}\,{\frac
{{\beta}^{3}}{{\gamma}^{2}}}+{ \frac
{128}{125}}\,{\varepsilon}^{2}\beta\right )y-\\
\\
-{\frac {553}{625}}\,\gamma{\varepsilon}^{4}\beta+{ \frac
{409}{1250}}\,{\varepsilon}^{3}{\beta}^{2}-{\frac {18}{625}}\,{
\frac {{\beta}^{4}\varepsilon}{{\gamma}^{2}}}+{\frac
{72}{625}}\,{\frac {{\varepsilon}^{2}{\beta }^{3}}{\gamma}}+{\frac
{589}{1250}}\,{\gamma}^{2}{\varepsilon}^{5}=0
\end{gathered}
\end{equation}

Assuming $\alpha=1$, $\beta=1$, $\gamma=1$ and $\varepsilon=2$
from \eqref{e:5.17} we have the equation in the form

\begin{equation}
\begin{gathered}
\label{e:5.17a}y_{{{\it zzz}}}+y_{{{\it zz}}}+y_{{z}}
+yy_{{z}}+{y}^{2}-{\frac {16}{5}}\,b\,y+{\frac
{751}{125}}\,y+{\frac {10589}{625}}=0
\end{gathered}
\end{equation}

Consider the case $B_{{2}}=B_{{2}}^{(1)}=0$ and
$A_{{2}}=A_{{2}}^{(1)}=-12\,\gamma$. We obtain

\begin{equation}
\begin{gathered}
\label{e:5.18}B_{{1}}=0,\,\,\quad\,\,A_{{1}}=24\,\gamma\,a+\frac{12}{5}(\beta-\varepsilon\,\gamma)
\end{gathered}
\end{equation}

\begin{equation}
\begin{gathered}
\label{e:5.19}A_{{0}}=8 \,\gamma\,b-{\frac
{12}{5}}\,\beta\,a+\frac{1}{25}\,{\frac {{\beta}^{2}}{\gamma}}
+{\frac
{23}{25}}\,\beta\,\varepsilon-\alpha-4\,\gamma\,{a}^{2}+{\frac {
12}{5}}\,\gamma\,a\varepsilon-{\frac
{24}{25}}\,\gamma\,{\varepsilon}^{2}
\end{gathered}
\end{equation}

\begin{equation}
\begin{gathered}
\label{e:5.20}C_{{0}}=\frac{4}{5}
\,\gamma\,\varepsilon\,{a}^{2}+\frac{4}{5}\,\gamma\,\varepsilon\,b-{\frac
{126}{125}}\,\gamma\,{\varepsilon}^{3}-\frac{4}{5}\,\beta\,{a}^{2
}-\frac{4}{5}\,\beta\,b-{\frac {3}{125}}\,{\frac
{\varepsilon\,{\beta}^{2}}{
\gamma}}+\\
\\
+{\frac
{128}{125}}\,{\varepsilon}^{2}\beta-\varepsilon\,\alpha+{\frac
{1}{125}} \,{\frac {{\beta}^{3}}{{\gamma}^{2}}}
\end{gathered}
\end{equation}

\begin{equation}
\begin{gathered}
\label{e:5.21}b={\frac {1}{100}}\,{\varepsilon}^{2}-{a}^{2}-{\frac
{1}{50}}\,{\frac { \beta\,\varepsilon}{\gamma}}+{\frac
{1}{100}}\,{\frac {{\beta}^{2}}{{ \gamma}^{2}}}
\end{gathered}
\end{equation}

\begin{equation}
\begin{gathered}
\label{e:5.22}C_{{1}}=\frac{1}{2}\,\varepsilon\,{\alpha}^{2}+\gamma\,{
\varepsilon}^{3}\alpha-{\varepsilon}^{2}\beta\,\alpha+{\frac
{409}{1250}}\,{\varepsilon}^{3}{ \beta}^{2}-{\frac
{18}{625}}\,{\frac {\varepsilon\,{\beta}^{4}}{{\gamma}^{2
}}}-\\
\\
-{\frac {553}{625}}\,\gamma\,{\varepsilon}^{4}\beta+{\frac
{72}{625}}\,{\frac {{\varepsilon}^{2}{\beta}^{ 3}}{\gamma}}+{\frac
{589}{1250}}\,{\gamma} ^{2}{\varepsilon}^{5}
\end{gathered}
\end{equation}

As a result of calculations we have the exact solitary waves in
the form that were found in \cite{9,16}

\begin{equation}
\begin{gathered}
\label{e:5.23}y(z)=-12\,\gamma\,{Y}^{2}+\left
(24\,\gamma\,a+{\frac {12}{5}}\,\beta- {\frac
{12}{5}}\,\varepsilon\,\gamma\right )Y-\\
\\
-{\frac {12}{5}}\,\beta\,a+{ \frac {3}{25}}\,{\frac
{{\beta}^{2}}{\gamma}}+{\frac {19}{25}}\,\beta
\,\varepsilon-\alpha-12\,\gamma\,{a}^{2}+{\frac
{12}{5}}\,\gamma\,a \varepsilon-{\frac
{22}{25}}\,\gamma\,{\varepsilon}^{2}
\end{gathered}
\end{equation}
where $Y(z)$ satisfies the equation

\begin{equation}
\begin{gathered}
\label{e:5.24}Y_{{z}}+{Y}^{2}-2\,aY-{\frac
{1}{100}}\,{\varepsilon}^{2}+{a}^{2}+{\frac {1}{50}}\,{\frac
{\beta\,\varepsilon}{\gamma}}-{\frac {1}{100}}\,{\frac {
{\beta}^{2}}{{\gamma}^{2}}}=0
\end{gathered}
\end{equation}
and is expressed by the function

\begin{equation}
\begin{gathered}
\label{e:5.25}Y(z)=a+\frac{1}{10}\,\left ({\frac
{\beta}{\gamma}}-\varepsilon\right )\tanh\left(\frac{1}{10}
\,\left ({\frac {\beta}{\gamma}}-\varepsilon\right )\left
(z-C_{{2}} \right )\right)
\end{gathered}
\end{equation}

Equation \eqref{e:3.28} can be presented in the form

\begin{equation}
\begin{gathered}
\label{e:5.26}\gamma\,y_{{{\it zzz}}}+\beta\,y_{{{\it
zz}}}+\alpha\, y_{{z}}+yy_{{z}}+\frac{1}{2}\,
\varepsilon\,{y}^{2}+\gamma\,{\varepsilon} ^{3}\alpha-{
\varepsilon}^{2}\beta\,\alpha+\\
\\
+\left ({\frac {126}{125}}\,\gamma\,{\varepsilon}^{3}+\frac{4}{5}
\,\beta\,{a}^{2}+\frac{4}{5}\,\beta\,b+{\frac {3}{125}}\,{\frac
{\varepsilon\,{\beta}^{2}}{\gamma}}-{\frac
{128}{125}}\,{\varepsilon}^{2}\beta\right)y-\\
\\
-\left(\varepsilon
\,\alpha+\frac{4}{5}\,\gamma\,\varepsilon\,{a}^{2}+\frac{4}{5}\,\gamma\,\varepsilon\,b+{
\frac {1}{125}}\,{\frac {{\beta}^{3}}{{\gamma}^{2}}}\right )y+\frac{1}{2}\,\varepsilon\,{\alpha}^{2}+\\
\\
+{\frac {409}{1250}}\,{\varepsilon}^{3}{\beta}^ {2}-{\frac
{18}{625}}\,{\frac
{\varepsilon\,{\beta}^{4}}{{\gamma}^{2}}}-{\frac
{553}{625}}\,\gamma\,{\varepsilon}^{4}\beta+{\frac
{589}{1250}}\,{\gamma}^{2} {\varepsilon}^{5}-{\frac
{72}{625}}\,{\frac {{\varepsilon}^{2}{\beta}^{3}}{ \gamma}}=0
\end{gathered}
\end{equation}

Assuming $\alpha=1$, $\beta=1$, $\gamma=1$ and $\varepsilon=2$
from \eqref{e:5.26} we have the equation in simple form

\begin{equation}
\begin{gathered}
\label{e:5.27}y_{{zzz}}+y_{{zz}}+y_{{z}}+y\,y_{{z}}+y^{2}-\frac{4}{5}{a}^{2}\,y-\frac{4}{5}b\,y\,+{\frac
{751}{125}}\,y+{\frac {5589}{625}}=0
\end{gathered}
\end{equation}

These exact solitary waves can be used to analyze nonlinear wave
processes at the description in a convective fluid.

\section{Exact periodic waves of the equation in a convective fluid}

Consider the equation for waves in a convective fluid again and
find the exact solutions in the form of periodic waves. We use the
following formula

\begin{equation}
\begin{gathered}
\label{e:6.1}y(z)=A_{{0}}+A_{{1}}\,Q+A_{{2}}Q^{2}+D_{{1}}Q_{{z}}+
B_{{1}}\frac{Q_{z}}{Q}+B_{2}\frac{Q_{{z}}^{2}}{Q^{2}}
\end{gathered}
\end{equation}
where $Q(z)$ is the general solution of equation \eqref{e:2.4}.

Substituting \eqref{e:6.1} into \eqref{e:3.28} we have $B_2^{(1)}
=-12\,\gamma, \,\,\, B_2^{(2)}=0$.

First of all let us consider the first case
$B_2=B_2^{(1)}=-12\gamma$.

We have

\begin{equation}
\begin{gathered}
\label{e:6.2}d=0,\,\,\quad\,c=0,\,\quad\,\,A_2=6\gamma
\end{gathered}
\end{equation}

\begin{equation}
\begin{gathered}
\label{e:6.3}D_{{1}}\,=\,\pm6\gamma,\,\,\quad\,\,B_{{1}}={\frac
{6}{5}}\,(\varepsilon\gamma-\beta),\,\,\quad\,\,A_{{1}}=9a\gamma
\,\mp\,\frac65 (\varepsilon\gamma-\beta)
\end{gathered}
\end{equation}

\begin{equation}
\begin{gathered}
\label{e:6.6}A_{{0}}=11b\gamma -\alpha +\frac{23}{25}
\beta\varepsilon +\frac{\beta^2}{25\gamma}-
\frac{24}{25}\gamma\varepsilon^2
\end{gathered}
\end{equation}

\begin{equation}
\begin{gathered}
\label{e:6.8}C_{{0}}={\frac
{\varepsilon\beta^2}{25\gamma}}-\varepsilon\alpha
-b\varepsilon\gamma -\frac{24}{25}
\gamma\varepsilon^3+\frac{23}{25}
\beta\varepsilon^2,\,\,\quad\,\,b=\frac{(b-\varepsilon
g)^2}{25\gamma^2}
\end{gathered}
\end{equation}

\begin{equation}
\begin{gathered}
\label{e:6.10}C_1 =\frac{409}{1250} \varepsilon^3\beta^2
-\varepsilon^2\beta\alpha-\frac{18}{625}
\frac{\varepsilon\beta^4}{\gamma^2}
+\frac{72}{625}\frac{\varepsilon^2\beta^3}{\gamma}
+\varepsilon^3\gamma\alpha+\\
\\+\frac12\varepsilon\alpha^2 +\frac{589}{1250} \varepsilon^5\gamma^2 -\frac{553}{625}
\varepsilon^4\gamma\beta.
\end{gathered}
\end{equation}

The exact periodic waves are determined by the formula

\begin{equation}
\begin{gathered}
\label{e:6.11}y(z)=\beta\,\varepsilon-\alpha-\gamma\,\varepsilon^2+
\left(\pm\,\frac65\,\beta \mp\,\frac65\,\varepsilon\,\gamma -3\gamma \,a\right)Q-\\
\\-6\,\gamma\, Q^2 \,\pm\,6\,\gamma \,Q_z +\frac65 \,(\varepsilon \,\gamma-
\beta)\,\frac{Q_z}{Q},
\end{gathered}
\end{equation}
where $Q(z)$ is the general solution of equation \eqref{e:2.4}.
Equation \eqref{e:3.28} in this case can be written in the form

\begin{equation}
\begin{gathered}
\label{e:6.12}\gamma\,y_{zzz}+\beta\,y_{zz}+
\alpha\,y_{z}+\frac{1}{2}\,\varepsilon\,y^{2}+y\,y_{z}+\varepsilon\,\alpha\,y\,+\gamma\,{\varepsilon}^{3}y-\beta\,{
\varepsilon}^{2}y-\\
\\
-{\varepsilon}^{2}\,\beta\,\alpha+{\varepsilon}^{3}\,\gamma\,\alpha+\frac{1}{2}\,\varepsilon\,\,{\alpha}^{2}+{\frac
{72}{625}}\, {\frac {{\varepsilon}^{2}{\beta}^{3}}{\gamma}}-{
\frac {553}{625}}\,{\varepsilon}^{4}\,\gamma\,\beta+\\
\\
+{\frac {409}{1250}}\,{\varepsilon }^{3}{\beta}^{2}+{\frac
{589}{1250}}\,{\varepsilon} ^{5}{\gamma}^{2}-{\frac
{18}{625}}\,{\frac {\varepsilon\,{\beta}^{4}}{{\gamma}^{2}} }=0.
\end{gathered}
\end{equation}

Assuming $\alpha=1$, $\beta=1$, $\gamma=1$ and $\varepsilon=1$ we
have simple form of equation \eqref{e:6.12}

\begin{equation}
\begin{gathered}
\label{e:6.13}y_{{zzz}}+y_{{zz}}+y_{{z}}+\frac{1}{2}\,y^{2}+y\,y_{z}+y+\frac12=0.
\end{gathered}
\end{equation}

Consider the second case $B_2=B_{{2}}^{(2)}=0$. We have

\begin{equation}
\begin{gathered}
\label{e:6.14}d=0,\,\,\quad\,\,B_{{1}}^{(1)}=\beta,\,\,\quad\,\,B_{{1}}^{{(2)}}=0
\end{gathered}
\end{equation}

Assuming $B_{1}=B_{1}^{(1)}=\beta$ we obtain

\begin{equation}
\begin{gathered}
\label{e:6.15}c=0,\,\quad\,\,A_{2}=-6,\,\gamma\,\quad\,D_{1}=\pm6\,\gamma
\end{gathered}
\end{equation}

\begin{equation}
\begin{gathered}
\label{e:6.16}A_{1}=\pm\frac{8}{3}\,\beta\,\mp2\,\varepsilon\,\gamma-3\,a\,\gamma,\,\,\quad\,\,
\beta=\frac{6}{11}\,\varepsilon\,\gamma
\end{gathered}
\end{equation}

\begin{equation}
\begin{gathered}
\label{e:6.17}A_{0}=-b\,\gamma-\frac{54}{121}\varepsilon^{2}\,\gamma-\alpha
\end{gathered}
\end{equation}

\begin{equation}
\begin{gathered}
\label{e:6.18}C_{0}=-\frac{54}{121}\,\varepsilon^{3}\,\gamma-\varepsilon\,
\alpha-b\,\varepsilon\,\gamma,\,\,\quad
\,b=\frac{1}{121}\,\varepsilon^{2}
\end{gathered}
\end{equation}

\begin{equation}
\begin{gathered}
\label{e:6.19}C_{1}=\frac{2989}{29282}\,\varepsilon^{5}\,\gamma^{2}+\frac{5}{11}\,
\varepsilon^{3}\,\gamma\,\alpha+\frac{1}{2}\varepsilon\,\alpha^{2}
\end{gathered}
\end{equation}

In this case we have special solution in the form

\begin{equation}
\begin{gathered}
\label{e:6.20}y(z)=-\frac{5}{11}\,\varepsilon^{2}\,\gamma-\alpha-\left
(\pm\frac{6}{11}\,\varepsilon\,\gamma+3\,a\,\gamma\,\right
)Q-6\,\gamma\,{Q}^{2}+\\
\\
\pm\,6\,\gamma\,Q_{{z}}+\frac{6}{11}\varepsilon\,\gamma\,\frac{Q_{{z}}}{Q}.
\end{gathered}
\end{equation}

Equation \eqref{e:5.8} for this case takes the form

\begin{equation}
\begin{gathered}
\label{e:6.21}\gamma\,y_{{{\it zzz}}}+{\frac
{6}{11}}\,\varepsilon\,\gamma\,y_{{{\it zz
}}}+\alpha\,y_{{z}}+\frac12\,\varepsilon\,{y}^{2}+y\,y_{{z}}+{\frac
{5}{11}}\,{\varepsilon}^{3}\,\gamma\,y+\varepsilon\,\alpha\,y+\\
\\
+{\frac {5}{11}}\,{\varepsilon}^{3}\gamma
\,\alpha+\frac12\,\varepsilon\,{\alpha}^{2}+{\frac
{2989}{29282}}\,{ \varepsilon}^{5}{\gamma}^{2}=0.
\end{gathered}
\end{equation}

In the case $\alpha=1$, $\beta=1$, $\gamma=1$ and $\varepsilon=1$
we have simple equation

\begin{equation}
\begin{gathered}
\label{e:6.22}y_{{{\it zzz}}}+y_{{{\it zz}}}+y_{{z}}+{\frac
{11}{12}}\,{y}^{2}+yy_{{ z}}+{\frac {1001}{216}}\,y+{\frac
{90695}{15552}}=0.
\end{gathered}
\end{equation}

Assuming $B_{2}=0$ and $B_{1}=0$ we have

\begin{equation}
\begin{gathered}
\label{e:6.23}A_{2}=-6\,\gamma,\,\,\quad\,D_{1}=\pm6\,\gamma,\,\,\quad\,\,A_{1}=\mp2\,
\varepsilon\,\gamma\pm\,2\,\beta-3\,a\,\gamma
\end{gathered}
\end{equation}

\begin{equation}
\begin{gathered}
\label{e:6.25}\beta=\varepsilon\,\gamma\,\,\quad\,\,A_{0}=-\alpha-b\,\gamma,\,\,\quad\,\,
C_{0}=-\varepsilon\,\,\alpha
\end{gathered}
\end{equation}

\begin{equation}
\begin{gathered}
\label{e:6.27}C_{1}=\frac{3}{2}\,\varepsilon\,\gamma^{2}\,a\,c-\frac{1}{2}\,b^{2}\,\varepsilon\,\gamma^{2}+
\frac{1}{2}\,\varepsilon\,\alpha^{2}-6\,\varepsilon\,\gamma^{2}\,d
\end{gathered}
\end{equation}

Special solution in this case takes the form

\begin{equation}
\begin{gathered}
\label{e:6.28}y(z)=-\gamma\,b-\alpha-3\,a\,\gamma\,Q-6\,\gamma\,{Q}^{2}\pm6\,\gamma\,Q_{{z}}
\end{gathered}
\end{equation}
where $Q(z)$ satisfies equation

\begin{equation}
\begin{gathered}
\label{e:6.29}Q_{{z}}^{2}-Q^{4}-a\,Q^{3}-b\,Q^{2}-c\,{Q}-d=0
\end{gathered}
\end{equation}

Equation \eqref{e:5.8} in this case can be written in the form

\begin{equation}
\begin{gathered}
\label{e:6.30}\gamma\,y_{{{\it
zzz}}}+\varepsilon\,\gamma\,y_{{{\it zz}}}+\alpha\,y_
{{z}}+\frac12\,\varepsilon\,{y}^{2}+y\,y_{{z}}+\varepsilon\,\alpha\,y+\frac32\,
\varepsilon\,{\gamma}^{2}\,a\,c+\\
\\
+\frac12\,\varepsilon\,{\alpha}^{2}-\frac12\,
\varepsilon\,{\gamma}^{2}\,{b}^{2}-6\,\varepsilon\,{\gamma}^{2}\,d=0.
\end{gathered}
\end{equation}

For $\alpha=1$, $\beta=1$, $\gamma=1$ and $\varepsilon=1$ we have
simple equation

\begin{equation}
\begin{gathered}
\label{e:6.31}y_{zzz}+y_{zz}+y_{z}+\frac12\,y^{2}+y\,y_{z}+y+\frac1
2-\frac12\,{b}^{2}-6\,d+\frac32\,ac=0
\end{gathered}
\end{equation}

Special solutions \eqref{e:6.11}, \eqref{e:6.20} and
\eqref{e:6.28} of equation \eqref{e:3.28} are new.

\section{Conclusion}

Let us emphasize in brief the results of this work. We have
presented new approach to look for exact solutions of nonlinear
differential equations. Two basic ideas were taken into account
for our method. The first idea was to use the simplest nonlinear
ordinary differential equation with known general solution for the
finding new special solutions. The second idea was to take all
possible singularities of solution of equation studied into
consideration. Using these ideas we have suggested new formulas to
search exact solutions of nonlinear differential equations in the
form of solitary and periodic waves.

Our approach was applied to look for exact solutions of the
Kuramoto -- Sivashinsky equation and the equation for description
nonlinear waves in a convective fluid. These equations  are
noninegrable equations but they are very popular at the
description of nonlinear waves in the nonlinear science. New exact
solutions in the form of solitary and periodic waves of these
nonlinear differential equations were found.

\section {Acknowledgments}

This work was supported by the International Science and
Technology Center under Project No 1379-2.

\end{document}